\documentclass[aps,amsmath,twocolumn,showpacs]{revtex4}
\begin{document}

\title{Matter-wave bright solitons of $^7$Li gas in an expulsive potential}
\author{ Sk. Golam Ali }
\author{ B. Talukdar}
\email{binoyt123@rediffmail.com}
\affiliation{Department of Physics, Visva-Bharati University,
Santiniketan 731235, India}

\begin{abstract}
A Lagrangian based method is used to derive an analytical model for studying the dynamics of matter-wave bright soliton created in a harmonic potential which is attractive in the transverse direction and expulsive in the longitudinal direction. By means of sech trial functions and a Ritz optimization procedure, evolution eqautions are constructed for width, amplitude and nonlinear frequency chirp of the propagating soliton of the atomic condensate. Our eqaution for the width is an exact agreement with that of Carr and Castin $\left[ Phys. Rev. A \,{\bf{66}}, 063602 (2002)\right]$, obtained by more detailed analysis. In agreement with the experiment of Paris group $\left[ Science\, \,{\bf{296}}, 1290 (2002)\right]$, the expulsive potential is found to cause the soliton to explode for $N|a_s|=0.98$, $N$ being the number of atoms in the condensate and $a_s$, the scattering length of the atom-atom interaction.
\end{abstract}
\maketitle
Solitons are localized waves that can propagate over long distances with neither attenuation nor change in shape.
Such waves are formed only when their dispersion is compensated by the nonlinear effects of the medium.
Solitons appear in many diverse physical situations including waves in shallow  water , pulse propagation 
in optical fibers and plasma waves. Since the experimental observation of Bose-Einstein condensates (BEC) in 1995, it was realized that we can also have matter-wave solitons \cite{1}. In this case, the nonlinearity is produced by binary atom-atom interactions leading to the mean field $U({\vec{ r}})=\frac{4\pi\hbar^2 a_s|\psi|^2}{m}$, where $a_s$ is the s-wave scattering length for the atom-atom scattering and m, the atomic mass. Here $\psi({\vec{ r}},t)$ is the wavefunction or order parameter of the condensate. For $a_s>0$ we get dark solitons while for $a_s<0$ we have bright solitons. A dark soliton is a notch on the BEC with a characteristic phase steep across it. On the other hand, bright soliton is a peak.
\par
Almost simultaneously two different groups, one at Ecole Normale Sup\'eriore in Paris \cite{2} and other at Rice University \cite{3}, produced bright 
solitons  from a trapped $^7$Li BEC in the internal atomic state ${\vert} F=1, m_F=1\rangle$ 
 by continuously tuning the scattering length from a positive to a negative value with the help of Feshbach 
resonance  \cite{4} induced by an applied magnetic field. In the Paris experiment, the spherical trapping of the BEC 
was adiabatically deformed into a cylindrical geometry and  the condensate was finally released into a horizontal 1D waveguide such that the resulting force on the atoms can be conveniently represented by $-m\omega_z^2z$, where
$\omega_z$ is the imaginary frequency along the z direction. Understandably, this imaginary frequency is due to an offset magnetic field that produces an expulsive harmonic potential $-\frac{1}{2}m\omega^2_zz^2$. The main effect of this expulsive term in the evolution equation is that the center of mass of the BEC accelerates along
 the longitudinal direction. Thus within the framework of mean-field approximation the Paris experiment can be modelled by a 3D Gross-Pitaevskii equation
\begin{equation}
\left[ i\hbar\frac{\partial }{\partial t }+\frac{\hbar^2}{2m}\nabla^2-V(\vec{r})-U(\vec {r})\right] \psi=0
\end{equation}
with
\begin{equation}
V(\vec{r})=\frac{1}{2}m\left[ \omega_\bot^2\left( x^2+y^2\right) -\omega_z^2z^2\right] ,
\end{equation} 
  the confining potential. Here $\omega_{\bot}$ is the frequency of the radial trapping and the wave function
$\psi(r,t)$ is normalized to the number of particles $N$ in the condensate such that 
\begin{equation}
\int\mid\psi\mid^2 d{\vec{r}}=N.
\end{equation} 
\par
In the case strong cylindric radial confinement it will be convenient to work with a quasi one-dimensional (Q1D) form of $(1)$ and thereby study the dynamics of matter-wave bright solitons created in a harmonic potential which  is attractive in the transverse direction but expulsive  in the longitudinal direction. We shall envisage 
such a study with particular emphasis on the role of expulsive potential in causing explosion as observed in the Paris experiment \cite{2}. \par 
There exists a number of detailed studies in respect of this. Remarkable among them are the works of Carr and Castin \cite{5} and of Salasnich \cite{6}. Our object in this work is to demonstrate how a simple analytical model  could be used to physically realize the problem of soliton explosion. We shall show that the critical number of atoms in the soliton just before explosion is in exact agreement with the numbers predicted by the sophisticated works in refs. 5 and 6 . To that end we first obtain Q1D equation from $(1)$. We write this equation in terms of dimensionless variables defined by
\begin{eqnarray}
\tau=\nu t,&\rho={r\over a_0},&
s={z\over a_0},\,\,\,\,
\psi(r, z,t)= {u(\rho,s,\tau)/ a_0^{3\over 2}}  \nonumber
\end{eqnarray}
with $a_{0}=\sqrt{\frac{\hbar}{m\omega_{\perp}}}$, the size of the ground state solution of the noninteracting GP equation. This gives
\begin{subequations}
\begin{equation} 
iu_{\tau}+{1\over 2}\nabla^2 u-{1\over 2}\left(\rho^2-\lambda_z^2 s^2\right)u-\frac{4\pi a_s}{a_0}|u|^2u=0 ,
\lambda_z=\frac{\omega_z}{\omega_{\perp}} .
\end{equation}
It is obvious that
\begin{equation}
\int|u|^2 d^3\rho = N .
\end{equation} 
\end{subequations}
Using a separable ansatz 
\begin{equation}
u(\rho,s,\tau)=\phi(\rho)\xi(s,\tau) 
\end{equation} 
$(4a)$ can be written in the form
\begin{eqnarray}
 {1\over {\xi}} \left(i\xi_{\tau}+{1\over 2}\xi_{2s}-{1\over 2}\lambda_z^2 s^2\xi\right)-\frac{4\pi a_s}{a_0}|\xi|^2|\phi|^2 
 \nonumber\\={1\over {\phi}}\left(-{1\over 2}\nabla^2_{\rho}\phi+{1\over 2}\rho^2\phi\right),
\end{eqnarray}
where  $\nabla^2_{\rho}$ stands for the Laplacian in the radial coordinate. In $(6)$ the subscripts on $\xi$ stands for
partial derivative with respect to that particular independent variable. More specifically,
$\xi_{2s}=\frac{\partial^2 \xi}{\partial s^2}$.  This equation shows that the presence of atom-atom interaction 
does not permit clearcut separation variables. However, the fourth term in equation $(6)$ is quite small such that $ \phi$ may be assumed to satisfy
\begin{equation}
-{1\over 2}\nabla^2_{\rho}\phi+{1\over 2}\rho^2\phi=\omega_{\rho}\phi
\end{equation}
with $\omega_{\rho}$  related to $\omega_{\perp} $  by a scale factor.
Equation $(7)$ represents the well-known eigenvalue problem for the two dimensional harmonic oscillator with the ground sate solution given by
\begin{equation}
{\phi_0}(\rho)={\rm e}^{-\rho^2/2}.\nonumber
\end{equation}
Thus $(6)$ can be written in the form
\begin{equation}
 i\xi_{\tau}+{1\over 2}\xi_{2s}+{1\over 2}\lambda_z^2 s^2\xi-\frac{4\pi a_s}{a_0}|\xi|^2|\phi|^2\xi
 ={\omega_{\rho}}\xi.
\end{equation}
The low-frequency vibration along the $z$ direction is quite unlikely to excite the two dimensional bosonic oscillator from its ground 
state. In view of this we  multiply $(8)$ by $\phi \phi^{\star}$ and integrate over the $\rho$ coordinate to get
\begin{equation}
 i\xi_{\tau}+{1\over 2}\xi_{2s}+{1\over 2}\lambda_z^2 s^2\xi-\frac{2\pi a_s}{a_0}|\xi|^2\xi
 ={\omega_{\rho}}\xi.
\end{equation}
Equation $(9)$ can be written in a more convenient form by using the change of variable 
\begin{equation}
\xi(s,\tau)=\chi(s,{\tau})e^{-i{\omega_{\rho}}\tau}.
\end{equation}
Using $(10)$ in $(9)$  we get
\begin{subequations}
\begin{equation}
 i\chi_{\tau}+{1\over 2}\chi_{2s}+{1\over 2}\lambda_z^2 s^2\chi-\frac{2\pi a_s}{a_0}|\chi|^2\chi=0
\end{equation}
with
\begin{equation}
{\int}^{+\infty}_{-\infty}|\chi|^2 ds=N/{\pi}.
\end{equation}
\end{subequations}
\par
The initial-boundary value problem in $(11a)$ can be converted to a variational problem in a rather straightforward manner. For example, it is easy to see that the action principle
\begin{equation}
\delta\int\int{\cal L}(\chi, \chi^\star, \chi_s, \chi_s^\star, \chi_{\tau},\chi_{\tau}^\star) ds \,d{\tau}
\end{equation} 
is equivalent to $(11a)$ with the Lagrangian density
\begin{equation}
{\cal L}={i\over 2} \left(\chi \chi_{\tau}^\star-\chi^\star \chi_{\tau} \right)-{1\over 2}\lambda_z^2 s^2\chi\chi^\star+\frac{\pi a_s}{a_0}
\chi^2{\chi^\star}^2+{1\over 2}\chi_s^\star\chi_s.
\end{equation} 
As a solution for $(11a)$ we introduce the trial function
\begin{equation}
\chi(s,\tau)=A(\tau)\text{sech}\left( s/a(\tau)\right) e^{ib(\tau)s^2} .
\end{equation} 
with a  complex amplitude $A(\tau)$. Here $a(\tau)$ stands for the 
width of the distribution and $b(\tau)$, the frequency chirp. In terms 
of trial function in $(14)$ we obtain a reduced variational problem.
\begin{equation}
\delta \int\left<{\cal L}\right> d{\tau}=0
\end{equation} 
with 
\begin{equation}
\left<{\cal L}\right> = \int^{+\infty}_{-\infty}{\cal L}_S \, ds .
\end{equation}  
Here ${\cal L}_S$ stands for the values of ${\cal L}$ when $(14)$ is substituted in $(13)$. We perform the       integration in $(16)$ to get 
\begin{eqnarray}
{\cal L}=ia\left( A A^{\star}_{t}- A^{\star} A_{t}\right) +\frac{\pi^2}{6}AA^{\star}b_ta^3-\frac{\pi^2}{12}\lambda_z^2a^3 AA^{\star}+\nonumber\\ \frac{4\pi a_s}{3a_0}aA^2{A^{\star}}^2
+\frac{AA^{\star}}{3a}+\frac{\pi^2}{3}b^2a^3AA^{\star}  .\,\,\,\,
\end{eqnarray} 
The vanishing conditions of $\frac{\delta\left<{\cal L}\right>}{\delta A}$,  
$\frac{\delta\left<{\cal L}\right>}{\delta A^\star}$, $\frac{\delta\left<{\cal L}\right>}{\delta a}$ and 
$\frac{\delta\left<{\cal L}\right>}{\delta b}$  yield
\begin{subequations}
\begin{eqnarray}
2iaA^{\star}_{t}+ia_tA^{\star}+\frac{\pi^2}{6}a^3b_tA^{\star}-\frac{\pi^2}{12}\lambda_z^2a^3A^{\star}+\nonumber\\
\frac{8\pi{a_s}}{3a_0}aA{A^{\star}}^2+\frac{A^{\star}}{3a}+\frac{\pi^2}{3}b^2a^3A^{\star}=0 \,\,\,  ,
\end{eqnarray} 
\begin{eqnarray}
2iaA_{t}+ia_tA-\frac{\pi^2}{6}a^3b_tA+\frac{\pi^2}{12}\lambda_z^2a^3A-\nonumber\\
\frac{8\pi{a_s}}{3a_0} aA^{\star}{A}^2-\frac{A}{3a}-\frac{\pi^2}{3}b^2a^3A=0 \,\, \, ,
\end{eqnarray} 
\begin{eqnarray}
i\left( AA^{\star}_{t}-A^{\star}A_{t}\right) +\frac{\pi^2}{12}b_ta^2AA^{\star}-\frac{\pi^2}{4}\lambda_z^2a^2AA^{\star}+ \nonumber\\
\frac{4\pi{a_s}}{3a_0} {A}^2{A^{\star}}^2-\frac{AA^{\star}}{3a^2}+\pi^2b^2a^2AA^{\star}=0 \,\, \,  ,
\end{eqnarray}
and 
\begin{eqnarray}
\frac{d}{dt}\left( a^3 AA^{\star}\right) -4ba^3AA^{\star}=0 \,\,\, .
\end{eqnarray}
\end{subequations}
Equations $(18a)$ and $(18b)$ can be combined to get 
\begin{equation}
a\vert A \vert^2= \,\,{\rm{Constant}} \,\,\, .
\end{equation} 
From the normalization of the trial function, the constant in $(19)$ can be identified with $N/{2\pi}$ such that
\begin{equation}
a\vert A \vert^2= \frac{N}{2\pi} \,\, \, .
\end{equation} 
Equations $(18a)$ - $(18c)$ and $(20)$ can be combined to write 
\begin{equation}
a^2b_\tau-\frac{1}{2}\lambda_z^2a^2-\frac{2}{\pi^2}\frac{Na_s}{a_0}\frac{1}{a}-\frac{2}{\pi^2a^2}+2a^2b^2=0  .
\end{equation} 
From $(18d)$ and $(20)$ we have 
\begin{equation}
b=\frac{1}{2a}\frac{da}{dt}  .
\end{equation} 
Using $(22)$ in $(21)$ we obtain a second-order differential equation for $a(\tau)$ given by
\begin{equation}
a_{2\tau}-\lambda_z^2a-\frac{4}{\pi^2}\frac{Na_s}{a_0}\frac{1}{a^2}-\frac{4}{\pi^2a^3}=0   .
\end{equation} 
Equation $(23)$ can be integrated to get
\begin{equation}
\frac{1}{2}a_{\tau}^2+V(a)=E
\end{equation} 
with 
\begin{equation}
V(a)=-\frac{1}{2}\lambda_z^2a^2-\frac{P}{a}+\frac{2}{\pi^2}\frac{1}{a^2}   .
\end{equation} 
Here $E$ is a constant of integration and $P=-\frac{4}{\pi^2}\frac{Na_s}{a_0}$.
\par
The extrema of the potential function in $(25)$ are determined from $\frac{dV(a)}{da}=0$. This gives
\begin{equation}
\pi^2\lambda_z^2a^4-\pi^2\vert P\vert a+4 =0 \,\,\, .
\end{equation} 
It is of interest to note that $(26)$ is an exact agreement with equation $(19)$ of Carr and Castin \cite{5} who studied the dynamics of matter-wave bright soliton with special attention on the variation of Gross-Pitaevskii 
energy functional. However, as opposed to the result in ref. 5, the width $a$ in $(26)$ is a function of time. This implies that $(26)$ holds good for any value of $\tau$. Whether these extrema are saddle points, minima, maxima is determined by
\begin{equation}
\frac{d^2V(a)}{da^2}=-\lambda_z^2-\frac{2\vert P\vert}{a^3}+\frac{12}{\pi^2a^4}\,\,.
\end{equation}  
In writing $(26)$ and $(27)$ we have introduced $P=-\vert P\vert$ to carry out the subsequent analysis by using  only numerical values of $a_s$. We note that for bright solitons the nonlinear term is attractive and the scattering length $a_s<0$. The critical point for collapse caused by the mean field, in the case $w_z^2<0$, 
for explosion due to expulsive potential may be obtained by simultaneously solving $(26)$ and $(27)$ for $a$ and 
 $\vert P\vert$ when $\frac{d^2V(a)}{da^2}=0$. We have 
\begin{equation}
a=\left( \frac{4}{3\pi^2\lambda_z^2}\right)^{\frac{1}{4}} \,\,\, {\rm{and}}\,\,\, \vert P\vert=\frac{16}{3\pi^2a}\,\,.
\end{equation} 

Equation in $(28)$ can be combined to get 
\begin{equation}
N|a_s|=\left(4/3\right)^{3/4}\pi^{1/2}\lambda_z^{1/2}a_0\,\, .
\end{equation} 
Using the experimental parameters $a_0=1.43  \mu\rm{m}$, $\lambda_z=\frac{70}{710}$ we have $N|a_s|=0.98$ for which
the expulsive potential causes the soliton to explode axially. This theoretical result is in agreement with the prediction of Khaykovich et al \cite{2} to within 10\%.  We, therefore,  conclude by noting that there are distinct advantages
to viewing nonlinear dynamics of bright solitons within the framework of a Lagrangian based approach.
\par
This  work forms the  part  of a Research Project F.10-10/2003(SR) supported  by the University Grants Commission, Govt. of India. One of the authors (SGA) is thankful to the UGC, Govt. of India for a Research Fellowship.

\end{document}